\documentclass{elsart}
\usepackage{natbib}
\usepackage{amssymb}

\begin{document}

\def\b{\begin{equation}}
\def\e{\end{equation}}
\def\l{\left}
\def\r{\right}

\begin{frontmatter}
\title{On the Non-occurrence of Type I X-ray Bursts from the Black Hole Candidates}

\author{Abhas Mitra}
\address { Theoretical Astrophysics Section, BARC, Mumbai,
India, Abhas.Mitra@mpi-hd.mpg.de, amitra@barc.ernet.in}
\date{}
\begin{abstract}
It   has been  justifiably questioned that if the Black Hole
Candidates (BHCs) have  a ``hard surface'' why Type I X-ray bursts
are not seen from them (Nayayan, R.,  Black Holes in Astrophysics,
New J. Phys., 7, 199-218, 2005). It is pointed out that a ``physical
surface'' need not always be ``hard'' and in could be ``gaseous'' in
case the compact object is sufficiently hot (Mitra, A., The day of
reckoning: the value of the integration constant in the vacuum
Schwarzschild solution, physics/0504076, p1-p6; BHs or Eternally
Collapsing Objects: A review of 90 years of misconceptions, in Focus
on Black Hole Research, Nova Sc. Pub., NY, p1-p94, 2006).
 Even if a ``hard surface'' would be there, presence of strong strong intrinsic magnetic
  field could inhibit Type I X-ray burst from a compact object
 as is the case for Her X-1. Thus, non-occurrence of Type I bursts actually rules out those alternatives of
 BHs which are either non-magnetized or {\em cold} and, hence, is no evidence for
 existence of Event Horizons (EHs). On the other hand, from the first principle, we again show
 that the BHCs being uncharged and having finite masses cannot be BHs, because uncharged BHs have a unique mass
 $M=0$. Thus  the previous results that the so-called BHCs are actually extremely
 {\em hot}, ultramagnetized,
 Magnetospheric Eternally Collapsing Objects (ECOs)
 ( Robertson, S.  \& Leiter, D., Astrophys. J. 565, 447-451, 2002; MECO Model of galactic BH Candidates and active galactic
 nuclei, in: New Developments in BLack Hole Research, NOva Sc. Pub., NY, p1-p44, 2006)
  rather than anything else get reconfirmed by non-occurrence of
 Type I X-ray bursts in BHCs.
\end{abstract}
\end{frontmatter}
\section{INTRODUCTION}  The concept of BHs arose as Schwarzschild, Hilbert and others tried to find the
spacetime geometry around a {\em MASSENPUNKT},  or a ``Point Mass''
having radius $R_0 =0$ \citep{mit1, mit2}:
\b
ds^2 =  \l(1 - {\alpha_0\over R}\r) dT^2
 -  \l({dR^2\over 1- {\alpha_0/ R}}\r) - R^2  d\Omega^2;~d\Omega^2 =d\theta^2 + \sin^2\theta d\phi^2
\e
Here $\alpha_0$ is an {\em undetermined  integration constant}.
Both the the interior and exterior spacetime of the BH is, however, represented by the so-called Eddington -Finkelstein metric
\b
ds^2 = \l(1-{\alpha_0\over R}\r) dT_*^2 \pm {2 \alpha_0\over R} dT_* dR
 - \l(1 +{\alpha_0\over R}\r) dR^2
+ R^2 d\Omega^2, \e \b T_* = T \pm \alpha_0  \ln\l({R-\alpha_0\over
\alpha_0}\r); \qquad dT_* = dT \pm {\alpha_0\over R-\alpha_0} dR \e
Let us recall now the fundamental property associated with any
curvilinear coordinate transformation: $\int \sqrt{-g} ~d^4x =
INVARIANT$ where $g$ is the determinant associated with the metric
coefficients. Then we can use this property for metrics (1) and (2):
$\int \sqrt{-g_*} ~dT_*~ dR~ d\theta~ d\phi = \int \sqrt{-g} ~dT ~dR
~d\theta ~ d\phi$. It can be easily verified that $ \sqrt{-g_*} =
\sqrt{-g} = R^2 \sin \theta $. Then, we will have, $\int R^2
~dT_*~dR~\sin\theta~d\theta~d\phi = \int R^2 ~dT
~dR~\sin\theta~d\theta~d\phi$. If we first carry  out the $\theta$
and $\phi$ integrations on both sides of this Eq., we will obtain
$\int R^2 ~dT_*~dR = \int R^2 ~dT ~ dR$. Then by using the form of
$dT_*$ from Eq.(3) in the above equation: \b \int R^2 ~dT~ dR \pm
\alpha_0 \int {R^2\over R -\alpha_0} ~dR ~dR = \int R^2 ~dT
~dR,~~or, \e \b \alpha_0 \int {R^2\over R -\alpha_0} ~ dR~ dR =0 \e
This equation can be satisfied only if $\alpha_0 = 2 M_0 \equiv
0$\citep{mit1}. Thus the uncharged Schwarzschild (actually Hilbert)
BHs have the unique mass $M=M_0 =0$. Hence, though, {\em
mathematically, BHs do exist}, their mass, $M_0\equiv 0$.  However,
mathematically, there could be charged BHs with $M\equiv Q$; but
astrophysical BHs are necessarily uncharged. And since the
(uncharged) BH paradigm was built by assuming $\alpha_0 = 2 M_0 >0$,
it collapses instantly. Moreover,  since the observed BHCs (or
anything else in the universe) have $M
>0$, they cannot be (uncharged) BHs!
It is important to note that Eq.(3) is obtained by integrating
Eq.(1) from $R=0$ to $R=R >R_0$, and, incase, we are considering  a
body with finite radius $R_0$, its {\em interior solution is not
covered by Eq.(1)}, and hence Eddington-Finkelstein metric is {\em
not valid in such a case}. Consequently, the procedure adopted here
would be irrelevant in such a case and  one would indeed have
$\alpha_0 \to \alpha = 2 M >0$. Mathematically while $M=
\int_0^{R_0} 4\pi R^2 ~\rho~dV >0$  for density $\rho >0$ and radius
$R_0 >0$, its value for a ``massenpunkt'', i.e., in the limit $R_0
\to 0$, is  $M_0 \equiv 0$.
\section{NON-OCCURRENCE of TYPE I BURSTS}
 As collapse of massive stars proceeds beyond the {\em cold}
Neutron Star (NS) stage having a small surface gravitational
redshift $z_0 = 0.2 -0.3$ and crosses the $(1+z_0)=\sqrt{3}$ limit,
it is believed to hurtle towards the BH stage having $z_0=\infty$.
If so, a very important physical phenomenon would occur. At high
$z_0$, photons/neutrinos generated within the body move along highly
curved paths and the chance of escaping decreases as $\sim
(1+z_0)^{-2}$. Density of radiation ($\rho_r$)  increases within the
collapsing body because of of both gravitational trapping and
matter-radiation interaction (diffusion) induced  trapping of
radiation.  It has been shown recently that because of such effects,
$\rho_r/\rho_0 \sim z_0 \gg 1$ \citep{mit3}.  Since BH formation
requires $z_0 \to \infty$, the collapsing object becomes almost a
pure ball of radiation {\em even before the formation of an EH}.
Note, in the usual collapse folklore, the Equation of State of the
collapsing object would remain practically unchanged and
$\rho_r/\rho_0 \ll 1$ before formation of any EH. Consequently, in
this folklore, one would promptly find a BH of the mass of the
original star core. But what actually happens is that as the trapped
radiation pressure would attain its corresponding Eddington value,
the outward radiation/ heat outflow pressure would dynamically stop
the collapse. The object then becomes a `` Relativistic Radiation
Pressure Supported Star'' (Mitra 2006b)  having a radius $R\approx 2
M$ and $z_0 \gg 1$. In a strict sense, the object is radiating and
contracting, and since this process becomes eternal, it is called an
``Eternally Collapsing Object'' (ECO). There is practically no
chance that the hot collapsing object would mysteriously turn into a
cold scalar field or a ``Dark Energy Star''. As, $z_0 \to \infty$
(BH), asymptotically, for an ECO, $M \to M_0 =0$.

It has been specifically shown that, the BH candidates have strong
intrinsic magnetic field by virtue of their physical surface which
in turn has shown that BHCs {\em do not have any Event Horizon}
\citep{rl1, rl2}. Accordingly, they are likely to be ECOs rather
than anything else. But Narayan (2005) and his coworkers have
repeatedly raised the valid question - if the BHCs have a
``surface'', why  they do not entertain Type I X-ray bursts. It is
forgotten here that, even if the BHCs would possess a real ``hard
surface''
 Type I burst activity
would be suppressed in  the presence of a strong surface magnetic
field because of (a) Modification of Scattering Cross-sections, and
(b) Strong magnetic channeling of the accretion flow on tiny
hot-spots. This is the reason that an accreting X-ray pulsar such as
Her X-1 does not show any Type I X-ray burst, and if the argument of
Narayan would be taken seriously, the pulsar in Her X-1 would have
an ``Event Horizon''.  Thus the ECOs  with strong intrinsic magnetic
fields, would not show any Type I burst activity. Hence the argument
of Narayan, at best, rules out the exotic {\em cold} BHCs and
further consolidates the idea of ECOS as the BHCs. In reality, the
stellar mass ECOs are extremely hot with local temperature exceeding
$200$ MeV and they could be in a state of Quark Gluon Plasma
\citep{mit4}. Thus although, ECOs will have a physical surface, they
have no ``hard surface''. Any nuclear matter accreted onto a MECO
may turn into QGP or get lost in the existing sea of QGP. The
question of any Type I X-ray burst would not arise in this scenario.

There could be circumstantial evidences that the BHCs are not BHs:
(1) Many BHCs do exhibit occurrence of ultra-relativistic jets, and,
though no EH has ever been detected, it was generally believed that
somehow {\em mysteriously} and self-contradictly, such jets must be
associated with EHs (when ``nothing can escape from an EH'').
However, the compact object in Cir X-1, has a physical surface and
has been found to launch an ultra relativistic jet with bulk
$\Gamma> 10$ \citep{fender}. Similarly, all ultrarelativistic jets
may actually be associated with compact objects having physical
surfaces. In general, presence of an intrinsic magnetic field may be
necessary for both launching and collimation of jets. In fact, we
know with certainty that there are objects without any EH but
intrinsic magnetic field; and often they do launch jets: e.g.,
protostellar clouds, stars, Neutron Stars etc. In contrast, not a
single case is known with certainty, where an object with an EH and
without any intrinsic magnetic field has ever launched a jet because
an EH has never been be detected! (2) As discussed \citep{vanderk},
the essential low freq. QPO behaviour of NSs and BHCs is the same.
The difference in the high freq. range may be due to stronger $B$
and $z$ of BHCs. (3) The large kick velocities associated with BHC
binaries cannot be explained if the BHCs are really BHs which are
formed by direct gravitational collapse\citep{vandenh} because
prompt formation of finite mass BH is supposed to be a quiet affair
accompanied by the prompt formation of an EH . On the other hand,
such large kick velocities can be easily understood if the BHCs are
objects with physical surfaces (but HOT) and formed in events
similar to powerful supernovae (Gamma Ray Bursts) whose
$\nu$-luminosity could be 100 times more than for typical SN events.

\end{document}